\documentstyle[10pt,aas2pp4]{article}
\slugcomment{Submitted to the J. of Korean Astron. Soc., Sep. 11, 1997.}

\def\spose#1{\hbox to 0pt{#1\hss}}
\def\lsim{\mathrel{\spose{\lower 3.0pt\hbox{$\mathchar"218$}}
     \raise 2.0pt\hbox{$\mathchar"13C$}}}
\def\gsim{\mathrel{\spose{\lower 3.0pt\hbox{$\mathchar"218$}}
     \raise 2.0pt\hbox{$\mathchar"13E$}}}
\def\msun{{\,M_\odot}}

\begin{document}
\twocolumn
\title{CORE-COLLAPSE TIMES OF TWO-COMPONENT STAR CLUSTERS}
\author{Sungsoo S. Kim\altaffilmark{1}}
\affil{Dept. of Physics, Space Sci. Lab.,\\
Korea Advanced Institute of Science \& Technology, Daejon, 305-701}
\authoremail{sskim@space.kaist.ac.kr}
\and
\author{Hyung Mok Lee}
\affil{Department of Earth Sciences, Pusan National University, Pusan, 609-735}
\authoremail{hmlee@uju.es.pusan.ac.kr}

\altaffiltext{1}{Most of his work has been done at his previous affiliation,
Institute for Basic Sciences, Pusan National University}

\begin{abstract}
We examine the corecollapse times of isolated,
two-mass-component star clusters using Fokker-Planck models.
With initial condition of Plummer models, we find that the corecollapse
times of clusters with $M_1/M_2 \gg 1$ are well correlated with\\
$(N_1/N_2)^{1/2}(m_1/m_2)^{2} t_{rh}$, where $M_1/M_2$ and $m_1/m_2$
are the light to heavy component total and individual mass ratios,
respectively, $N_1/N_2$ is the number ratio, and $t_{rh}$ is the initial
half-mass relaxation time scale.
We also find two-component cluster parameters that best match 
multi-component (thus more realistic) clusters with power-law mass functions.
\end{abstract}


\section{INTRODUCTION}

The course of dynamical evolution of pre- and postcollapse globular
clusters is determined by many factors such as initial mass function,
nature and efficiency of energy generation mechanisms, tidal cut-off,
anisotropy of velocity distribution, initial population of binaries,
and stellar evolution.
There have been many efforts in developing more and more complex cluster models
including such factors, making analysis and interpretation rather difficult.
To study the dynamical evolution of globular clusters more realistically,
among others, Chernoff \& Weinberg (1990) included the effects of stellar
evolution, Lee, Falhman, \& Richer (1991) used multi-component models,
Takahashi (1995) includeded an anisotropic velocity distribution.

However, studying simpler models could be more instructive in identifying
important physical processes governing the evolution.  Kim, Lee, \& Goodman
(1997; hereafter KLG) studied on the postcollapse evolution of cluster variables
and the gravitational oscillation using two-component Fokker-Planck models.
In this paper, as a supplementary study to KLG, we present a fitting formula
for the corecollapse times of two-component models and compare the results of
two- and multi-component models.  As in KLG, here both tidal-capture
binary heating and tree-body binary heating are included, and clusters are
assumed to be isotropic and isolated (no tidal cutoff).  For the methods that
we are using here and the benefits of studying simpler models (two-component
models), readers are referred to KLG and references therein.

Corecollapse times of two-component clusters were presented by Inagaki \&
Wiyanto (1984), Inagaki (1985), and Lee (1995) among others.  These papers
calculated corecollapse times as a function of $M_2/M_1$, the
ratio of total masses of heavy component to light component, and found that
the ratio of corecollapse time to initial half-mass relaxation
time, $t_{cc}/t_{rh}$, has a minimum value at $M_2/M_1 \sim 0.1$.
However, the paramter $M_2/M_1$ may be divided into $m_2/m_1$, the individual
mass ratio, and $N_2/N_1$, the number ratio.  In the present paper, we
calculate the corecollapse times of two-component models as a fuction of
more complete two-copmonent cluster parameters, and find a fitting formula
between them.  However, the clusters studied here are restricted to those with
$M_1 \gg M_2$ as in KLG.

On the other hand, it would be helpful in interpreting the results of
two-component models if the similarities and discrepancies between the results
of two- and multi-component models are well known.  In the present paper,
we also compare two-component models to 11-component models, and thus provide
a way to extrapolate to more realistic cluster the results of two-component
clusters such as those in KLG.

\section{CORE-COLLAPSE TIMES}

To calculate the corecollapse times of two-component star clusters, we have
performed total 11 runs of direct numerical integration of the orbit-averaged
Fokker-Planck equation with a local approximation.  The code used here is
descended from Cohn (1980).  Parameters of our two-component runs are shown
in Table
\ref{table:A}.  This set of parameters has been chosen such that it
provides all possible combinations of parameters $M$, $N$, $m_2/m_1$, and
$N_1/N_2$, where $M$ is the cluster mass and $N$ is the total number.
Note that in all our runs, the total mass of heavy component, $M_2$, is
negligible compared to the total mass of light component, $M_1$, and thus
$m_1 \approx M/N_1$.  The initial density and velocity profiles are given
by Plummer models with $v_{c1}/v_{c2}=1$ and $\rho_{c1}/\rho_{c2}=M_1/M_2$,
where $v_c^2$ is the three-dimensional core velocity dispersion, and $\rho_c$
is the core density.

\begin{deluxetable}{lcrrrrcr}
\tablecolumns{11}
\tablewidth{0pt}
\tablecaption{Parameters and Core-Collapse Times of Two-Component Models
\label{table:A}}
\tablehead{
\colhead{Run} &
\colhead{$m_2 \over m_1$} &
\colhead{$N_1 \over N_2$} &
\colhead{$M$} &
\colhead{$N$} &
\colhead{$m_2$} &
\colhead{$t_{cc}$} &
\colhead{$t_{cc} / t_{rh}$} \\
\multicolumn{3}{c}{} &
\colhead{($\msun$)} &
\colhead{} &
\colhead{($\msun$)} &
\colhead{($10^{10}$yr)} &
\colhead{}
}
\startdata
baab  &2 &100 &       $10^5$ &141457 &                 1.4 &1.27 & 12.42\nl
caab  &3 &100 &       $10^5$ &210125 &                 1.4 &0.93 &  6.34\nl
faab  &4 &100 &       $10^5$ &277473 &                 1.4 &0.61 &  3.23\nl
cdab  &3 & 30 &       $10^5$ &201299 &                 1.4 &0.56 &  3.97\nl
cbab  &3 &300 &       $10^5$ &212871 &                 1.4 &1.62 & 10.92\nl
caab1 &3 &100 &       $10^5$ & 70042 &        $3\times$1.4 &0.35 &  6.47\nl
caab2 &3 &100 &       $10^5$ &630374 &${1\over3}\times$1.4 &2.52 &  6.28\nl
baab3 &2 &100 &       $10^5$ &212185 &${2\over3}\times$1.4 &1.80 & 12.17\nl
faab3 &4 &100 &       $10^5$ &208104 &${4\over3}\times$1.4 &0.47 &  3.23\nl
caeb  &3 &100 &$3\times10^4$ & 63037 &                 1.4 &0.57 &  6.34\nl
cabb  &3 &100 &$3\times10^5$ &630374 &                 1.4 &1.49 &  6.43\nl
\tablecomments{The initial half-mass radii $r_{h}$ of these runs
are all 5 pc.}
\enddata
\end{deluxetable}

\begin{figure}[t]
\plotone{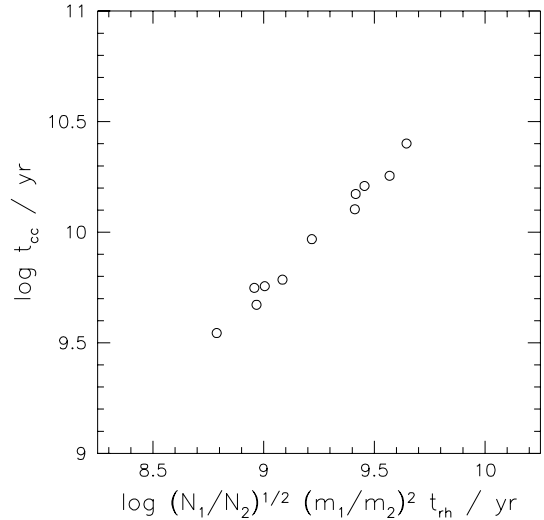}
\caption{\label{fig:tcc}Corecollapse times of runs in Group A.  $N_5 \equiv
N/10^5$ and $M_5 \equiv M/10^5 \msun$.}
\end{figure}

Corecollapse times of our runs are shown in Table \ref{table:A} in units of
$10^{10}\,{\rm yr}$ and $t_{rh}$.  We empirically found that $t_{cc}$ can
be fitted by the following formula:
\begin{eqnarray}
\label{tcc}
	t_{cc} \approx 4.2 \times 10^9{\rm yr}
			 \left ( {N_1 \over N_2} \right )^{1/2}
			 \left ( {m_1 \over m_2} \right )^2 & \nonumber \\
			 N_5 M_5^{-1/2}
			 \left ({r_{h} \over 5\, {\rm pc}} \right)^{3/2}, &
\end{eqnarray}
where $N_5 \equiv N/10^5$, $M_5 \equiv M/10^5 \msun$, and $r_h$ is the
initial half-mass radius.
Each $t_{cc}$ value is plotted over the righthand side of the above equation
in Figure \ref{fig:tcc}, which shows a good X-Y correlation.
Equation (\ref{tcc}) is to be compared with the standard half-mass relaxation
time scale,
\begin{equation}
\label{trh}
	t_{rh} \equiv {v_m^2 \over \langle v^2_{\parallel} \rangle_{v=v_m}}
		=     {M^{1/2}r_h^{3/2} \over 6.7 G^{1/2} m \ln 0.4N},
\end{equation}
where $v_m$ is the root-mean-square three-dimensional velocity of the whole
cluster and $\langle v^2_{\parallel} \rangle_{v=v_m}$ is the average change of
$v^2_m$ in parallel component to initial $v_m$ per unit time.

Isolated single-mass clusters with initial condition of Plummer models
collapse at 15.4 $t_{rh}$ (Cohn 1980), where $t_{rh}$ is the half-mass
relaxation time scale and does not vary much until the corecollapse takes
place.  However, the ratios of the time required for corecollapse $t_{cc}$
to $t_{rh}$ and core relaxation time scale $t_{rc}$ strongly depend on the
density and velocity profiles.  Quinlan (1996) found that for single-mass
clusters, $t_{cc}$ varies much less when expressed in units of $t_{rc}$
divided by a dimensionless measure of the temperature gradient in the core.
Although in single-mass clusters the velocity profile (as well as other
physical parameters) evolves by the two-body relaxation, two- or
multi-component clusters have another driving force: the equipartition.

Both mass segregation and equipartition are envolved in determination of
the time to corecollapse, and this complexity makes the theoretical
interpretation of the above correlation between $t_{cc}$ and cluster
parameters quite difficult.  Here we suggest the following
analysis as one way to explain this correlation.

The actual duration of corecollapse is very small compared to the time
to corecollapse from the begining of cluster's evolution.  Instead,
clusters spend most of their precollapse phases under mass segregation
process and approach to the onset of homologous phase of corecollapse.
If a considerable degree of equipartition is accomplished in the precollapse
phase as in all of our two-component models, the time to the onset of
corecollapse will be determined by how fast the light component gains
the energy from the heavy component via equipartition.  Thus one may
define the precollapse time scale of two-component clusters $t_{r2}$
as following:
\begin{equation}
\label{tr2}
	t_{r2} \equiv {v^2_{m1} \over \langle v^2_{\parallel 1}
			\rangle_2},
\end{equation}
where $\langle v^2_{\parallel 1}\rangle_2$ is the velocity dispersion
change of the light component via interactions with heavy component.  Using
the standard expression for the average velocity dispersion change per unit
time, one has
\begin{equation}
\label{Dv2}
	\langle v^2_{\parallel 1} \rangle_2 \propto
		{G^2m_2\rho_{m2} \over v_{m2}},
\end{equation}
where the heavy component mean density $\rho_{m2}$ is
proportional to $M_2/r_h^3$ and the Coulomb logarithm has been omitted.
It is also assumed that $v_{m1} \sim v_{m2}$.
Spitzer (1969, 1987) showed that for a two-component cluster
of polytropic index $n$ between 3 and 5 with $M_1 \gg M_2$ and a Maxwellian
velocity distribution in a parabolic potential well, the minimum degree of the
global equipartition is a function of cluster's parameters such that
\begin{equation}
\label{min_equi}
	\left. {m_2v^2_{m2} \over m_1v^2_{m1}} \right |_{\rm min}
		\propto \left ( {N_2\over N_1}\right )^{2/3}
		        \left ( {m_2\over m_1}\right )^{5/3}.
\end{equation}

With the minimum value of equation (\ref{min_equi}) and assumptions that
$M_1 \gg M_2$ and $v^2_{m1} \sim GM/r_h$, equation (\ref{tr2}) now becomes
\begin{eqnarray}
\label{tr2a}
	t_{r2} & \propto & \left ( {N_1 \over N_2} \right )^{2/3}
			   \left ( {m_1 \over m_2} \right )^{5/3}
			   N M^{-1/2} r_{h}^{3/2}  \nonumber \\
	       & \propto & \left ( {N_1 \over N_2} \right )^{2/3}
			   \left ( {m_1 \over m_2} \right )^{5/3} t_{rh}.
\end{eqnarray}
The above is in the same form as equation (\ref{tcc}) with only small
discrepancies in the exponents.  Although equation (\ref{min_equi})
has been used for derivation of the above equation, we find
that the degrees of equipartition in the precollapse phases of our
two-component runs do not directly correlate with the minimum values of
equation (\ref{min_equi}).  In fact, exact equipartition is usually not
accomplished even when the value of equation (\ref{min_equi}) is less than
unity, because as mass segregation of heavy component progresses, interactions
between heavy and light components occur less.  Thus equation
(\ref{min_equi}) should be regarded as a degree of tendancy to equipartition
and it is this tendancy that $t_{r2}$ requires in its definition.

While Quinlan (1996) introduced a
temperature gradient in the core in a derivative form to explain a huge
variation in $t_{cc}$ for clusters with different initial profiles, here
we introduced both density and velocity gradients of heavy component
naturally into the time scale by considering global equipartition.

\section{COMPARISON WITH MULTI-\\
COMPONENT CLUSTERS}

Clusters have continuous mass functions.  However, mass functions are
usually realized with discrete mass components in numerical calculations.
Scientists found that 10 to 20 components are enough to represent continuous
mass functions for Fokker-Planck models, and such numerical representation
for a given mass function is quite straightforward for these multi-component
clusters:  there is only a question of choice of each component's mass bin
and a representative value.  However, when the number of components is reduced
to 2 for the sake of analytical simplification, such choice is not so simple
because dynamically important mass and corresponding number of stars may
be different from simple mean mass and total number of a certain mass range.
Therefore two-component cluster parameters (such as $m_2/m_1$, $N_1/N_2$,
and $N$) that well represent a continuous mass function should be numerically
found through comparisons of the evolution of two- and multi-component
clusters.

\begin{deluxetable}{lcrrcccc}
\tablecolumns{8}
\tablewidth{0pt}
\tablecaption{Parameters and Results of Multi-Component Models
\label{table:B1}}
\tablehead{
\multicolumn{5}{c}{} &
\multicolumn{3}{c}{Values at $t=10^{11} \, {\rm yr}$} \\ \cline{6-8}
\colhead{Run} &
\colhead{$x$} &
\colhead{$M$} &
\colhead{$N$} &
\colhead{$t_{cc}$} &
\colhead{$\rho_c$} &
\colhead{$v_c$} &
\colhead{$r_h$} \\
\multicolumn{2}{c}{} &
\colhead{($\msun$)} &
\colhead{} &
\colhead{($10^{10}$yr)} &
\colhead{($\msun \, {\rm pc}^{-3}$)} &
\colhead{(${\rm km} \, {\rm s}^{-1}$)} &
\colhead{(${\rm pc}$)}
}
\startdata
A2 & 1.00& $        10^5$ &  402857 & 0.528 & $7.63\times 10^4$ & 2.79 & 32.8\nl
B2 & 1.35& $        10^5$ &  509201 & 0.584 & $8.13\times 10^4$ & 2.75 & 27.7\nl
C2 & 1.50& $        10^5$ &  552232 & 0.621 & $8.19\times 10^4$ & 2.73 & 26.2\nl
B1 & 1.35& $3\times 10^4$ &  152760 & 0.356 & $1.71\times 10^3$ & 1.30 & 38.5\nl
B3 & 1.35& $3\times 10^5$ & 1527603 & 0.931 & $2.58\times 10^6$ & 5.44 & 20.7\nl
\tablecomments{The initial half-mass radii $r_{h}$ of these runs
are all 5 pc.}
\enddata
\end{deluxetable}

In this section we will compare our multi-component models with
the two-component models in KLG varying $M$ and the mass function of the
multi-component models.  Cluster parameters of our multi-component models
are given in Table \ref{table:B1}.  The initial density and velocity
profiles are given by Plummer models.  The initial half-mass radii of all
multi- and two-component models are 5 pc.  The number of component
is 11 and we adopt a power-law mass function:
\begin{equation}
	N(m) dm \propto m^{-(x+1)} dm,
\end{equation}
where $x$ is the mass spectral index and the Salpeter mass function has
$x=1.35$.  For a bin $i$ with boundaries $m_{ia}$ and $m_{ib}$, the total
mass in the bin is obtained by
\begin{equation}
	M_i=\int^{m_{ib}}_{m_{ia}} m N(m) dm.
\end{equation}
Then the number of stars in the bin is $N_i=M_i/m_i$, where $m_i$ is
the representative mass of each bin.  The main-sequence star mass range
was selected to be $0.08\msun - 0.8\msun$.  Following Sigurdsson \& Phinney
(1995), the stars of initial mass $m_i$ between $0.8 \msun$ and $4.7 \msun$
were assumed to have evolved to white dwarfs of mass $0.58+0.22
\times(m_{\rm MS}-1.0) \msun$, where $m_{\rm MS}$ is the main-sequence mass,
while stars of $m_{\rm MS}$ between $4.7 \msun$ and $8.0 \msun$ were
assumed to disrupt completely.  The stars heavier than $8.0 \msun$ but
lighter than $15.0 \msun$ were assumed to become neutron stars of mass
$1.4 \msun$.  Neutron stars are born with a kick velocity due to an
asymmetric explosion, and they are ejected from the cluster if the kick
velocity is greater than the escape velocity of the cluster.  However, we
assumed that all neutron stars remain in the cluster, because the retention
rate of neutron stars are not well known and the precise realization of
real clusters is not our goal in this study.  The mass range, representative
mass, and number of stars of each component is shown in Table \ref{table:B2}.
Our multi-component models include both three-body binary heating and
tidal-capture binary heating, but we find that the postcollapse phases of all
our runs are driven by three-body binary heating.

\begin{deluxetable}{rcl@{ -- }lcccccccc}
\tablecolumns{12}
\tablewidth{0pt}
\tablecaption{Mass Spectra of Multi-Component Models
\label{table:B2}}
\tablehead{
\colhead{Bin} &
\colhead{$m_i$} &
\multicolumn{2}{c}{Mass Range} &
\multicolumn{2}{c}{$x=1.00$} &
\colhead{} &
\multicolumn{2}{c}{$x=1.35$} &
\colhead{} &
\multicolumn{2}{c}{$x=1.50$} \\ \cline{5-6} \cline{8-9} \cline{11-12}
\colhead{} &
\colhead{($\msun$)} &
\multicolumn{2}{c}{($\msun$)} &
\colhead{$N_i$} &
\colhead{$m_i \cdot N_i$} &
\colhead{} &
\colhead{$N_i$} &
\colhead{$m_i \cdot N_i$} &
\colhead{} &
\colhead{$N_i$} &
\colhead{$m_i \cdot N_i$}
}
\startdata
 1& 0.1& 0.08& 0.15& 1.00000& 1.00000&& 1.00000& 1.00000&& 1.00000 & 1.00000 \nl
 2& 0.2& 0.15& 0.25& 0.40631& 0.81262&& 0.33263& 0.66526&& 0.30517 & 0.61034 \nl
 3& 0.3& 0.25& 0.35& 0.17842& 0.53526&& 0.12584& 0.37752&& 0.10826 & 0.32478 \nl
 4& 0.4& 0.35& 0.45& 0.09995& 0.39980&& 0.06359& 0.25436&& 0.05223 & 0.20892 \nl
 5& 0.5& 0.45& 0.55& 0.06385& 0.31925&& 0.03753& 0.18765&& 0.02985 & 0.14925 \nl
 6& 0.6& 0.55& 0.65& 0.11712& 0.70272&& 0.05766& 0.34596&& 0.04264 & 0.25584 \nl
 7& 0.7& 0.65& 0.80& 0.09010& 0.63070&& 0.04104& 0.28728&& 0.02945 & 0.20615 \nl
 8& 0.9& 0.80& 1.0 & 0.02428& 0.21852&& 0.00822& 0.07398&& 0.00516 & 0.04644 \nl
 9& 1.1& 1.0 & 1.2 & 0.01286& 0.14146&& 0.00389& 0.04279&& 0.00232 & 0.02552 \nl
10& 1.3& 1.2 & 1.4 & 0.00774& 0.10062&& 0.00215& 0.02795&& 0.00124 & 0.01612 \nl
11& 1.4& 1.4 & 1.4 & 0.00913& 0.12782&& 0.00182& 0.02548&& 0.00091 & 0.01274 \nl
\tablecomments{$N_i$ and $m_i \cdot N_i$ are normalized with bin 1 values.}
\enddata
\end{deluxetable}

We find a two-component model which best describes a given multi-component
model by comparing the values of cluster variables $\rho_c$, $v_c$,
$r_h$ at $t=10^{11}\,\rm{yr}$, and $t_{cc}$.  An epoch of $10^{11}\,\rm{yr}$
has been selected as in KLG because by that time, our runs have reached
self-similar expansion phase.  With two-component models, KLG found the
following numerical values:
\begin{mathletters}
\label{scale_sim}
\footnotesize
\begin{eqnarray}
        \rho_c & \simeq & 4.5 \times 10^5 \, \msun / {\rm pc^3} \, \,
                \left ({m_2 \over m_1} \right )^{-10/3}
                N_5^{10/3} t_{11}^{-2.0}; \\
        v_c   & \simeq & 3.8 \, {\rm km/s} \, \,
                \left ({m_2 \over m_1} \right )^{-1/2}
                N_5^{1/3} M_5^{1/3} t_{11}^{-0.32}; \\
        r_c   & \simeq & 0.042 \, {\rm pc} \, \,
                \left ({m_2 \over m_1} \right )^{7/6}
                N_5^{-4/3} M_5^{1/3} t_{11}^{0.65}; \\
        r_h   & \simeq & 35 \, {\rm pc} \, \,
                N_5^{-2/3} M_5^{1/3} t_{11}^{0.65},
\end{eqnarray}
\end{mathletters}
where $N_5 \equiv N/10^5$, $M_5 \equiv M/10^5\msun$, and $t_{11} \equiv t/
10^{11}\,{\rm yr}$.
On the other hand, the numerical values from our multi-component models
are given in Table \ref{table:B1}.  There are four two-component parameters
to be determined for a given multi-component model, $m_2/m_1$, $N_1/N_2$,
$N$, and $M$.  However, since cluster variables at a certain
epoch in the postcollapse phase are independent of $N_1/N_2$ as in
equation (\ref{scale_sim}), $N_1/N_2$ has to be determined from the
corecollapse time, equation (\ref{tcc}).  Then the rest three parameters,
$m_2/m_1$, $N$, and $M$, may be determined from equation (\ref{scale_sim}).
This method will be called Method A, and parameters obtained in this way
are given in Table \ref{table:B3}.

\begin{deluxetable}{crrrrrrrrrr}
\tablecolumns{11}
\tablewidth{0pt}
\tablecaption{Best Matching Two-Component Cluster Parameters
\label{table:B3}}
\tablehead{
\colhead{} &
\multicolumn{4}{c}{Method A} &
\colhead{} &
\multicolumn{5}{c}{Method B} \\ \cline{2-5} \cline{7-11}
\colhead{Run} &
\colhead{$m_2 \over m_1$} &
\colhead{$N_1 \over N_2$} &
\colhead{$N_5$} &
\colhead{$M_5$} &
\colhead{} &
\colhead{$m_2 \over m_1$} &
\colhead{$N_1 \over N_2$} &
\colhead{$N_5$} &
\colhead{$M_5$} &
\colhead{$N_2$}
}
\startdata
A2 & 1.78 & 13.1 & 1.05 & 0.90 && 1.70 & 10.9 & 1.00 & 0.82 &  8400 \nl
B2 & 2.33 & 28.5 & 1.40 & 0.97 && 2.17 & 21.3 & 1.30 & 0.84 &  5800 \nl
C2 & 2.56 & 39.4 & 1.54 & 0.99 && 2.36 & 28.4 & 1.41 & 0.84 &  4800 \nl
B1 & 2.74 & 53.9 & 0.51 & 0.35 && 2.37 & 30.0 & 0.44 & 0.26 &  1400 \nl
B3 & 2.05 & 18.1 & 3.47 & 2.49 && 2.00 & 16.3 & 3.38 & 2.36 & 20000 \nl
\tablecomments{$N_5 \equiv N/10^5$ and $M_5 \equiv M/10^5 \msun$.
$N_2$ has been approximately calculated by $N/(N_1/N_2+1)$ and
has only two significant digits.}
\enddata
\end{deluxetable}
 
With Method A, our multi-component model B2 is best described by a
two-component model with $m_2/m_1=2.3$, $N_1/N_2=29$, $N=1.4\times 10^5$,
and $M=0.97\times 10^5 \msun$.  Note that
with these parameters, $m_2 \simeq 1.6 \msun$, which is little higher than
the mass of the heaviest component of our multi-component models, $1.4 \msun$.
Neutron stars play an important role in dynamical evolution of globular
clusters: a considerable fraction of dynamical binary formation (as apposed to
primordial binaries) envolves neutron stars.  For this reason, in two-component
clusters, the heavy component is often targeted for neutron stars and the
light component is for main-sequence stars.  Thus in finding the best matching
two-component parameters, it could be more meaningful if $m_2$ is set to
$1.4 \msun$.  With a restriction of
$m_2 = 1.4 \msun$, now the number of variables in equation (\ref{scale_sim})
required for determination of two-component cluster parameters is reduced to
two.  Since $\rho_c$ and $r_h$ are two cluster variables that represent
the status of the core and envelope, respectively, we use these variables
along with $m_2=1.4\msun$ and equation (\ref{tcc}) for our second method
(Method B) to find the best matching two-component model (see Table
\ref{table:B3}).

With Method B, model B2 is now best described by a two-component model
with $m_2/m_1=2.2$, $N_1/N_2=21$, $N=1.3\times 10^5$, and $M=0.84 \times
10^5 \msun$.  Note that with these parameters, $N_2 = 4890$ and
this value is about the same with the number of stars in the heaviest four bins
(bins only for degenerate stars) of model B2 ($\sum_{i=8,11} N_i=5176$).
This may imply that the epoch of corecollapse is mainly determined by
the number of stars above the turnoff mass.  This fact also holds for
other runs with different $M$ and $x$.

For clusters with $N \propto M$ (as for our multi-component clusters B1, B2,
and B3), equation (\ref{scale_sim}) may be written as $\rho_c \propto M^{10/3}$,
$v_c \propto M^{2/3}$, and $r_h \propto M^{-1/3}$.  From runs B1, B2, and B3
in Table \ref{table:B1}, $\rho_c$, $v_c$, and $r_h$ are found to be proportional
to $M^{3.18}$, $M^{0.622}$, and $M^{-0.269}$, respectively.  The absolute
values of these exponents are little smaller than equation (\ref{scale_sim}).
However, since the discrepancies are not so significant, we conclude that the
evolution aspects of the postcollapse multi-component clusters are still
well predictable from the numerical and analytical results of two-component
clusters.  On the other hand, cluster variable values at
$t=10^{11} \,{\rm yr}$ show relatively small x dependence.

The results from Method B in Table \ref{table:B3} indicate that
multi-component clusters may be described by two-component clusters
with masses 15 to 20 \% less and with $m_1$ near the turnoff mass.  Of course,
this $m_1$ is dependent on $x$ such that clusters with steeper mass function
are matched by two-component clusters with smaller $m_1$.  However,
interestingly, lighter multi-component clusters {\it with the same $x$} also
require smaller $m_1$.  This comes from the fact that $N \propto M$ holds
for runs B1, B2, and B3, while not for their matching two-component clusters
by Method B ($N \propto M^{0.92}$).  For Method A, the best matching
two-component clusters of runs B1, B2, and B3 show $N \propto M^{0.98}$, which
results in nearly the same $m_1$.  Thus we conclude that the above difference
in $m_1$ values by Method B for clusters with the same $x$ stems from the
restriction, $m_2 = 1.4 \msun$.

\begin{figure}[t]
\plotone{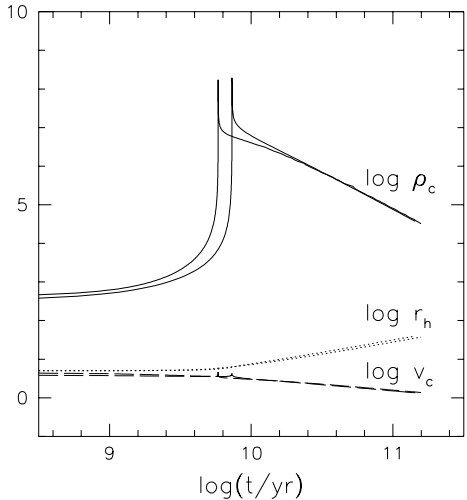}
\caption{\label{fig:2multi}Comparison of the evolution of multi-component
model (run B2; thick lines) and best-matching two-component model by Method B
($m_2/m_1=2.2$, $N_1/N_2=21$, $N=1.3\times 10^5$, and
$M=0.84 \times 10^5 \msun$; thin lines).  The units of $\rho_c$, $v_c$,
and $r_h$ are $\msun \,{\rm pc^{-3}}$, ${\rm km \, s^{-1}}$, and ${\rm pc}$,
respectively.}
\end{figure}

The evolution of the two-component model with the above parameters is
plotted in Figure \ref{fig:2multi} as well as that of model B2.
Cluster variables $\rho_c$, $v_c$ and $r_h$ of two runs well coincide.
Only the corecollapse times show a small discrepancy.  This is partly
because of the dispersion of $t_{cc}$ from the fitting formula,
equation (\ref{tcc}), and partly because of the small $N_1/N_2$ value:
equation (\ref{tcc}) is to be used for clusters with $M_1 \gg M_2$.

\section{SUMMARY}

We have investigated the evolution of isolated two-component
clusters with initial condition of Plummer models.  The corecollapse time
$t_{cc}$ showed a good correlation with a parameter
$(N_1/N_2)^{1/2}(m_1/m_2)^2 t_{rh}$.  To explain this correlation, a new
time scale for the precollapse evolution of two-component clusters,
$t_{r2}\equiv (N_1/N_2)^{2/3}(m_1/m_2)^{5/3} t_{rh}$ have been introduced
using Spitzer's (1969, 1987) global equipartition analysis.

We also found two-component clusters which best match with
our multi-component clusters with power-law mass functions.
For example, the evolution of 11-component cluster with a Salpeter mass
function and $M=10^5 \msun$ was well described by a two-component
cluster with $m_2/m_1=2.2$, $N_1/N_2=21$, $N=1.3\times 10^5$, and
$M=0.84 \times 10^5 \msun$.  Furthermore, it has been found that the best
matching two-component cluster has $N_2$ very close to the number of stars
heavier than turnoff mass of the multi-component cluster.

\acknowledgements
S. S. K. thanks Chang Won Lee and Jung-Sook Park for obtaining old and rare
papers.  This research was supported in part by the Matching Fund Programs
of Research Institute for Basic Sciences, Pusan National University,
RIBS-PNU-96-501, and in part by Basic Science Researech Institute Program
to Pusan National University under grant No. 95-2413.


\end{document}